**An investigation of reconfigurable magneto-mechanical metamaterials**

Russell Galea[1], Pierre-Sandre Farrugia[1,2], Krzysztof K. Dudek[3], Louis Zammit Mangion[4], Joseph N. Grima[5] and Ruben Gatt[1,6,*]

[1] *Metamaterials Unit, Faculty of Science, University of Malta, Msida MSD 2080, Malta*

[2] *Department of Geosciences, Faculty of Science, University of Malta, Msida MSD 2080, Malta*

[3] *Institute of Physics, University of Zielona Gora, ul. Szafrana 4a, Zielona Gora 65-069, Poland*

[4] *Department of Physics, Faculty of Science, University of Malta, Msida MSD 2080, Malta*

[5] *Department of Chemistry, Faculty of Science, University of Malta, Msida MSD 2080, Malta*

[6] *Centre for Molecular Medicine and Biobanking, University of Malta, Msida, MSD 2080, Malta*

*Corresponding author contact information*:

Prof R Gatt: ruben.gatt@um.edu.mt (ORCID-iD: 0000-0002-1950-743X)

Mobile: +356 99 211 477

*Other email contact information*:

Mr R Galea: russell.galea.15@um.edu.mt (ORCID-iD: 0000-0003-1624-5291)

Prof P S Farrugia: pierre-sandre.farrugia@um.edu.mt, (ORCID-iD: 0000-0002-4108-965X)

Dr K K Dudek: k.dudek@if.uz.zgora.pl (ORCID-iD: 0000-0002-9570-6713)

Dr Louis Zammit Mangion: louis.zammit-mangion@um.edu.mt, (ORCID-iD: 0000-0002-9507-842X)

Prof J N Grima: joseph.grima@um.edu.mt (ORCID-iD: 0000-0001-5108-6551)

**Abstract:**

In this work, an experimental investigation of an accordion-like system with embedded magnetic inclusion is carried out. It is shown that the way that this structure deforms when subjected to an external magnetic field depends on the magnetic moment of the embedded magnets and the length of the arms. Stacking of the accordion-like system leads to the formation of hexagonal honeycombs that can have a negative, zero, and positive Poisson's ratio. Variation in the configuration attained depends on the relative positioning of the magnetic inclusion and the applied magnetic field. In particular, one of the hexagonal honeycomb arrangements was able to switch between a conventional and a re-entrant configuration upon the reversal of the external magnetic field. For all structures considered, the dimensions can be controlled through the external magnetic field allowing for a high degree of turnability. Furthermore, their behaviour can be altered in real time. The practical implications of the results are of interest since they indicate that these structures can be adopted in numerous applications, such in the design of scaffoldings for deployable structures, actuators, variable pored sieves, and sound proofing systems.

**Keywords:** Magneto-mechanical metamaterial; auxetic; magnetic inclusions; shape-programmable materials;

## 1. Introduction

Shape-programmable materials (SPMs) are a class of active materials whose geometry and dimensions are dependent on specific stimuli, such as heat,[1–5] light,[6,7] chemicals,[8–13] and pressure[14,15] as well as electrical[16,17] and magnetic fields.[18–33] The introduction of SPMs gives rise to materials which are more responsive and adaptable, resulting in a superior mechanical functionality when compared to traditional mechanisms, especially when considering miniature devices.[34] In fact, SPMs have been utilised in soft robotics, as well as employed as actuators, deployable devices, and biomedical devices due to a more precise and finer control.[15] One promising class of shape-programmable materials comprises the magnetically actuated materials, as their shape can be manipulated not only through the magnitude of the stimuli but also through the direction and spatial gradient of the magnetic field allowing the creation of complex configurations.[35–41]

The ductility of SPMs can be further extended by incorporating functional elements from systems exhibiting a negative Poisson's ratio, i.e., that contract laterally when uniaxially compressed and *vice versa*. This unusual behaviour generally stems from the specific geometry and the mode of deformation. Auxetic systems, as these structures are called,[42] have attracted increasing attention since the late 1980s.[43–70] The reason for this stems from the fact that they have improved functionalities as compared to their conventional counterparts. These include, amongst others, enhanced indentation resistance,[45,71–73] increased shear stiffness,[45,74] and fracture toughness,[45,75] as well as vibration control.[74,76,77] For this reason they have been suggested for a multitude of applications such as, impact loading,[78–83] medical dilator,[84] and sensors.[74,85]

The concept of incorporating auxetic characteristics in SPMs has already been explored in a number of works experimental and numerical studies. Tipton et al.[86] designed a magneto-elastic metamaterials by embedding magnets within an elastomers having periodic circular holes that is akin to the auxetic system proposed by Bertoldi et al.[87] When subject to an external magnetic field the structure adopted what can be termed as a twist-buckle deformation that mimics that of the corresponding auxetic system under compression. A system based on the rotating squares mechanism with the inclusion of permanent magnets was studied by Slesarenko.[38] This investigation indicated that some mechanical properties, such as buckling strain and post-buckling stiffness, could be finely tuned through the strategic placement of the magnets. Scarpa and co-workers[88,89] showed how ferromagnetic particles seeding of an auxetic polyurethane foam resulted in superior acoustic absorption properties while allowing for the tuning of the peak acoustic absorption coefficient by using the external magnetic fields. Negative stiffness was also reported in the work of Dudek et al.[90] where magnetic inclusions were added to a structure based on the double arrow system. In this case, the magnitude of the stiffness could be adjusted by varying the orientation of magnetic inclusions and their magnetic moment. Other auxetic designs that could lend themselves to have magnetic inclusion, thus allowing them to reconfigure under the action of an external magnetic field, were also presented by Grima et al.[39] Numerical studies have also been key in helping in the understanding of the mechanisms behind auxetics having magnetic inclusions. In fact as early as 2007, Dudek et al.[91] performed simulations, analysing an auxetic ferrogel having magnetic inclusions.

More recently, Galea et al.[33] proposed an accordion-like structure with embedded magnets which could rapidly respond to an external magnetic field by reconfiguring. It was shown that the response was dependent on the strength for the external magnetic field, allowing for the dimensions of the structure to be altered. Theoretical and numerical studies revealed how the stacking of the basic accordion-like structure could lead to hexagonal honeycombs having a negative, zero, and positive Poisson's ratio. In this context, the work presented here complements and extends on this initial study so as to investigate experimentally the dependence of the response of these structures on the magnetic moment of the embedded magnets and the distance between the magnets using experimental methods.

## 2. Materials and methods

*2.1 System considered*

The basic system that will be considered can be described as having a nonmagnetic backbone in the form of an accordion-like foldable structure as shown in Figure 1b. Magnetic inclusions, consisting of permanent magnets, are inserted in the rigid arms (bars) of the backbone with the orientation being such that the internal magnetic forces allow the structure to attain an equilibrium position in the shape of an accordion. Once an external magnetic field is applied on the composite structure, the arms will fold. These can get closer together, farther away, or even flip the folding depending on the direction of the magnetic field. The reason for this is that the magnetic inclusion will tend to align with the external magnetic field thus inducing a rotation in the nonmagnetic arms. If the external magnetic field is removed, the internal forces of the system will induce it to return to its equilibrium position.

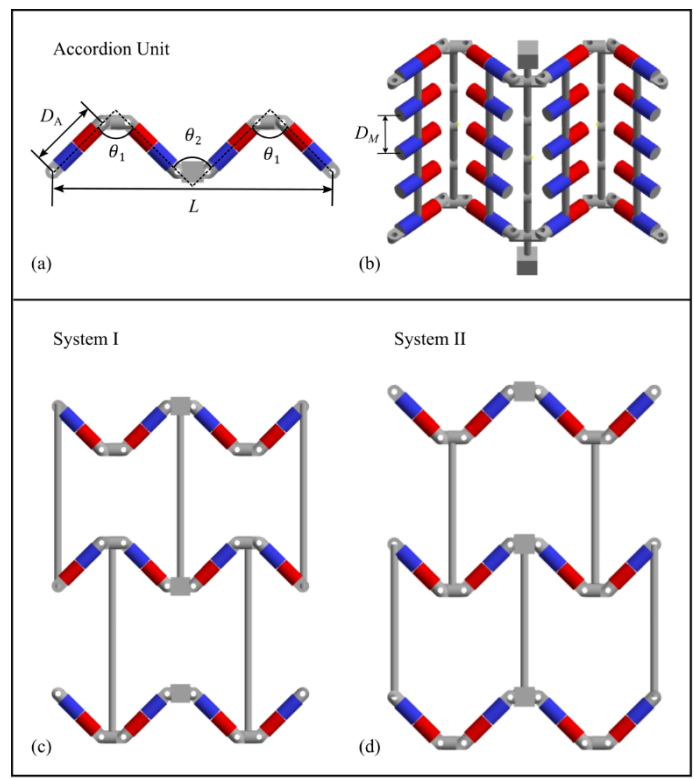

**Figure 1.** (a) Top and (b) tilted view of the accordion unit depicting the parameters considered in this study. (c-d) Top view of System I and System II arrangements indicating the orientation of the magnetic inclusions.

The accordion-like structure is characterised by two composite angles, labelled $\theta_1$ and $\theta_2$, as well as three lengths denoted by $L$, $D_A$, and $D_M$ (Figure 1a-b). Due to practical considerations in the physical prototypes a connection bar was introduced in order to join the two adjacent arms of the accordion-like structure. Each arm was then attached to the connection bar using a copper wire (having a diameter of 0.2 mm). This introduced two extra degrees of freedom. Thus, in this simple configuration, the $\theta_1$ and $\theta_2$ represent the smallest angle that successive arms make with each other as indicated in Figure 1a. In an ideal model, $\theta_1$ and $\theta_2$ would be equal. However, $\theta_1$ is untethered on one side which can induce it to exhibit edge effects whereas $\theta_2$ is constrained from both ends. For this reason the two angles can attain different values and hence need to be considered separately.

The length $L$ gives the length between two corresponding arms while $D_A$ gives the length of an arm. On the other hand, $D_M$ represents the distance between the magnets in the out of plane direction. In order to simplify the analysis $D_A$ and $D_M$ were taken to have the same magnitude and are subsequently referred to as $D$.

The basic configuration can then be stacked in different ways. Referring to Figure 1c connecting alternating corners in such a way so as to retain the same polarity of the magnets leads to a system that, under the action of an external magnetic field, attains a shape that is akin to the semi-re-entrant honeycomb,[92] which is well known to have a zero Poisson's ratio. This system will be referred to as System I. On the other hand, connecting the corners with successive layers of the accordion-like structure being sifted results in a structure that under the action of an external magnetic field will take the form of a re-entrant or conventional honeycomb system. These structures are characterised by a negative and positive Poisson's ratio respectively and this configuration will be referred to as System II.

*2.2 Experimental Procedure*

The accordion-like system was initially designed in Autodesk Inventor Professional 2022. This was then exported so as to be printed with an SLA 3D printer (Formlabs, Form 3) having a resolution of 25 μm in the *xy* plane and a layer height of 50 μm. Formlabs Tough 2000 Resin was used for the purpose. The prints were then cleaned in Formlabs Form Wash using

Isopropyl alcohol (IPA) for a total of 20 minutes. Following this the structure was cured in the Formlabs Form Cure at 60 °C for 1 hour.

In total six different structures were designed and built. Four of the structures had the parameter $D$ set to 8.5 mm with the magnetic inclusion holders being modified to house two to five cylindrical neodymium magnets each having a height and a diameter equal to 1 mm. These structures were referred to as M$i$, where $i$ represents the number of magnets. Experimental testing indicated that the M$i$ had a dipole moment of 1.82 mA m$^{-2}$, 2.73 mA m$^{-2}$, 3.64 mA m$^{-2}$, and 4.55 mA m$^{-2}$, for $i = 2,\ldots,5$ respectively. The other two structures, referred to as D2 and D3, were designed to house five magnets while $D$ was set to 12 and 17 mm. These were used to study the effect of varying $D$. The relative orientation of successive magnetic inclusions is indicated in Figure 1a-b.

The prototypes thus produced were placed between two duly calibrated electromagnets that were able to generate a uniform magnetic field between them (Figure 2a (i)). They were set up in such a way that the gravitational force had a minimal effect on the hinging of the structure as shown in Figure 2a (ii). The deformation of the structures when subjected to a magnetic field could then be monitored using an imaging camera (Daheng MER2-630-60U3M) mounted with a lens having a focal length of 1.4 (get cameras LCM-5MP-08MM-F1.4-1.5-ND1) stably fixed and levelled vertically above the system (Figure 2a (i)).

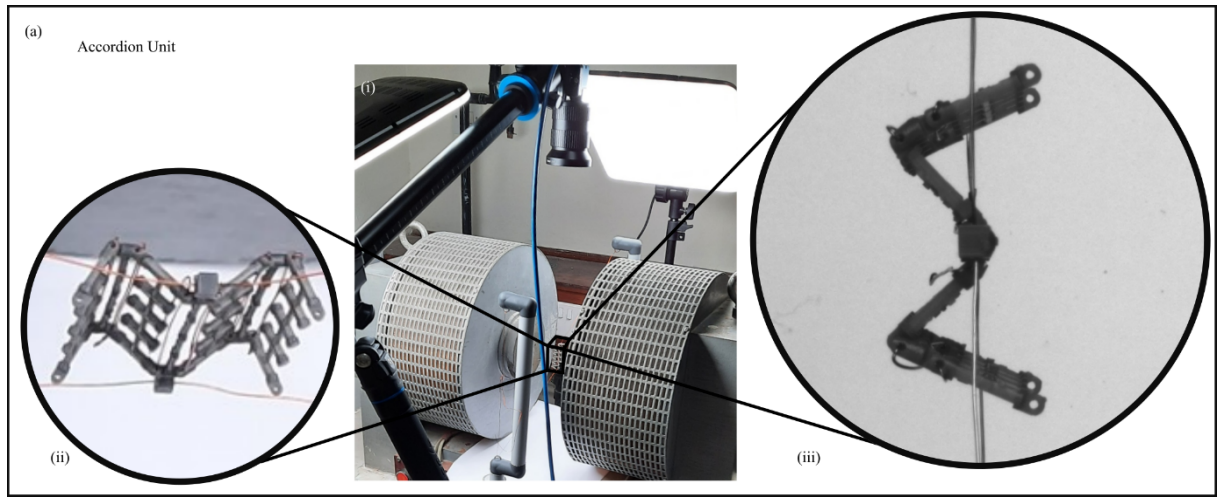

**Figure 2.** (a)(i) A depiction of the setup used in the experiments including a (ii) sideway image and (iii) the image recorded by the camera for the experiment.

The extension/contraction of the accordion-like systems was then studied using an external magnetic field having values ranging between 4.0 and 43.4 mT. In a typical experiment the electromagnets were switched on and the accordion-like system was allowed to attain the new equilibrium position. At this point the electromagnets were switched off and with the deformation being recorded by the camera. This was repeated ten times for each external magnetic field used. A python script was used to determine automatically $\theta_1$, $\theta_2$, and $L$ from the images obtained. To do so, reference points were selected manually for each image. $\theta_1$ and $\theta_2$ were in practice determined by extrapolating the arms to a point connecting them. On the other hand, $L$ was determined by measuring the distance between the external arms. Once the variables were calculated, the percentage change as compared to their values at the lowest external magnetic field (namely 4.0 mT) was determined. The use of the relative change in $\theta_1$ and $\theta_2$, and $L$ allowed for a better comparison of the deformation between the structures. A similar process was then applied for Systems I and II.

## 3. Results and Discussion

The influence of the magnetic moment on the behaviour of the structure was studied first. Its importance stems from the fact that it establishes the internal repulsive forces present within the system that maintain the stability of the structure. It also dictates the system's interaction with the external magnetic field. The results can pave the way to the introduction of electromagnets in place of the permanent magnets that are able to exert a varying and controllable internal magnetic field. In doing so, it should be possible to control the equilibrium position of the structure. Furthermore, it would be possible to vary the conformation of the structure in real time allowing the system to attain a desired morphology by applying different magnetic fields in different regions or positions.

In order to study the effect of the magnetic moment of the magnetic inclusions on the deformation of the structure M$i$ (where $i = 2,…,5$ and represents the number of magnetic inclusions) were subjected to different external magnetic fields. The results, which are shown in Figure 3, indicate that structure M2 had the smallest relative changes in $\theta_1$, $\theta_2$, and $L$, possibly due to the fact that the relatively low magnetic moment did not create sufficient torque to overcome the friction at the hinges. On the other hand, it can be noted that the changes in the angles $\theta_1$ and $\theta_2$ for all the other structures are basically the same. This suggests that once the friction at the hinges is overcome, the angles will increase to the same relative extent irrespective of the internal magnetic moment. Interestingly, this is not reflected in the change in $L$. There are various reasons for this. One factor is that the reference equilibrium value of $\theta_1$ and $\theta_2$ differed according to the magnitude of the magnetic moment. Furthermore, the connecting bar provided additional degrees of freedom that altered the simple dependence between $L$ and angles as suggested from Figure 1. Even so, at relatively low magnetic fields, the value of $L$ for M3, M4, and M5 are very similar. They start to diverge at around an external magnetic field of 15 mT. Remarkably, the smallest variation was that for M3 while the largest was for M4 with M5 being somewhere in between. This would suggest that there could be a reversal of the increasing of $L$ with increasing internal magnetic moment.

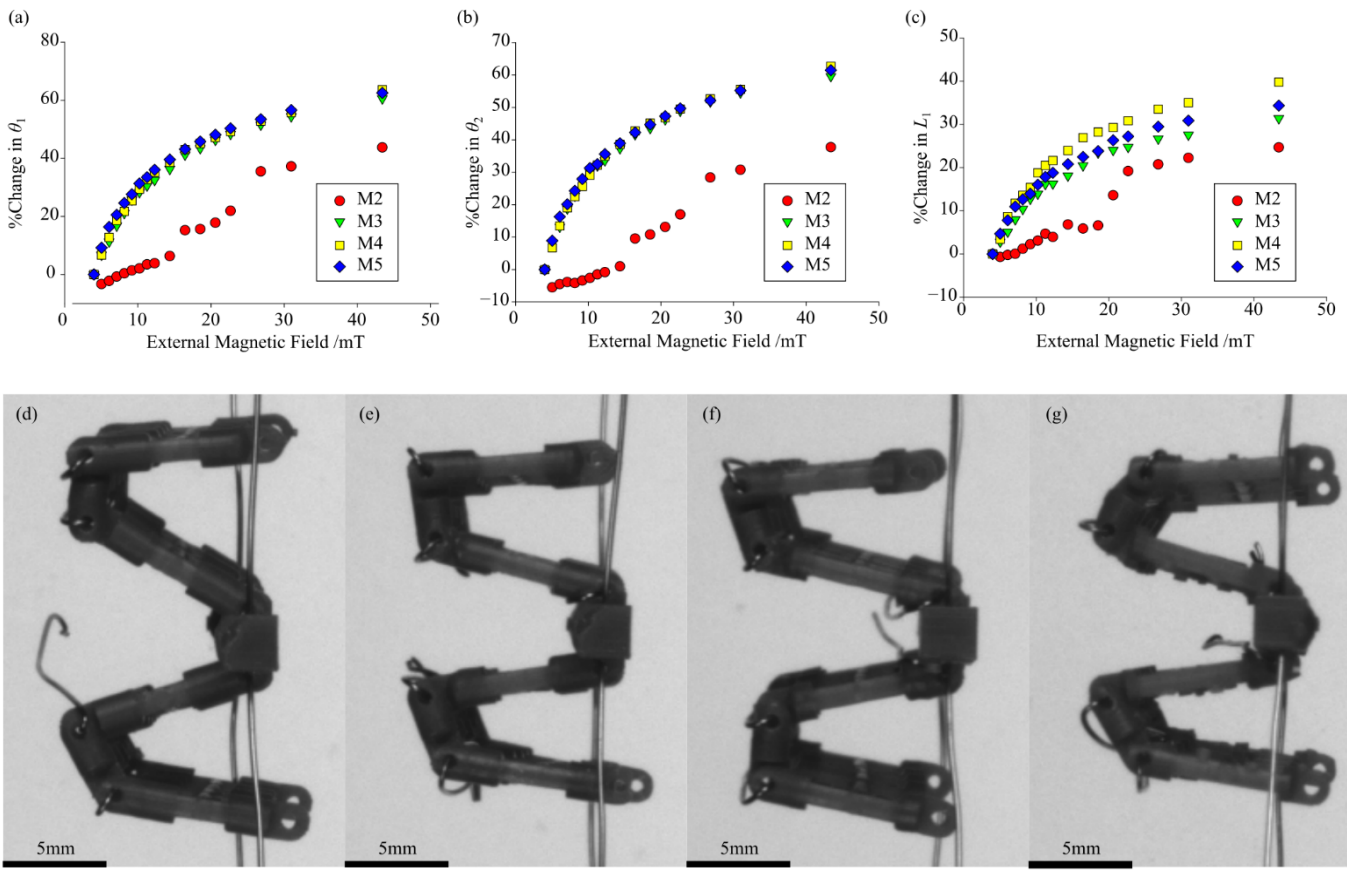

**Figure 3.** (a-c) Graphs representing the percentage change in $\theta_1$, $\theta_2$, and $L$ respectively against the applied external magnetic field for structures M$i$, where $i = 2,…,5$. (d-g) Images taken during the experiment when an external magnetic field of 43.4 mT was applied for structures M$i$, where $i = 2,…,5$, respectively.

The changes induced by varying the parameter $D$ were studied next. This parameter dictates the moments induced about the hinges due to the repulsive internal forces that act as restoring forces as well as those induced by the external magnetic field. The results are meant to shed light on how the behaviour of the system will change when the structure scaled up or down in order to accommodate a desired application and how the geometry affects the equilibrium configuration. Understanding the interplay of forces and moments acting on the structure is fundamental in order to design both miniature and larger devices based on the same principles presented here.

In order to study the effect of the parameter $D$ on the deformation of the structure the behaviour of M5 was compared with that of D2 and D3 when these were subjected to different magnetic fields. Each of these structures has five embedded magnets but the value of $D$ was set to 8.5, 12, and 17 mm respectively. The results, shown in Figure 4, indicate that the behaviour of M5 and D2 are similar but differ slightly from those of D3, particularly for the relative change in $\theta_1$ and $\theta_2$. In this case, it is unlikely that the difference in behaviour is due to the friction in the hinges. As a matter of fact, the longer arms of D3 should have allowed it to exert a larger torque at the joints. However, by the same token, a small change in $\theta_1$ and $\theta_2$ would have moved the embedded magnetics of D3 farther away (or closer) than in the case of M5 and D2.

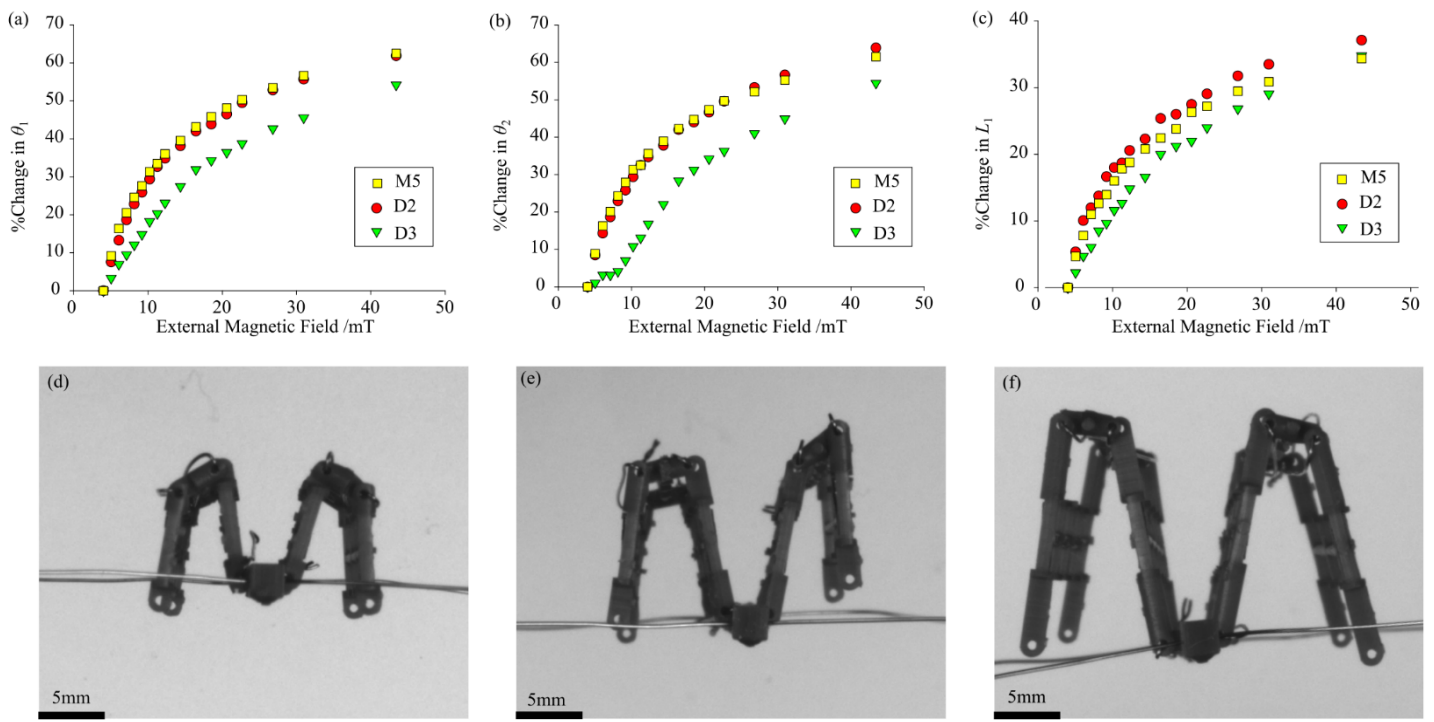

**Figure 4.** (a-c) Graphs representing the percentage change in $\theta_1$, $\theta_2$, and $L$ respectively against the applied external magnetic field for structures M5, D2, and D3. (d-g) Images taken during the experiment when the external magnetic field was set to 43.4 mT for structures M5, D2, and D3 respectively.

It is worth noting that the results for $L$ are not too different from one another. Yet there does not seem to be a discernible pattern, for the change in $L$ with increasing $D$. Once again, this could be partially explained by the fact that the initial values of $\theta_1$ and $\theta_2$ were different for the three structures. In addition, the introduction of a connecting bar added further degrees of freedom to the motion. Geometric effects relating the angles with $L$ as well as inducing different internal forces between the magnets could also have played a role.

As described in the methodology, these accordion-like systems may also be used to construct three-dimensional honeycomb systems. The behaviour of System I under the action of an increasing external magnetic field is depicted in Figure 5. The structure configured itself as a semi re-entrant hexagonal honeycomb under the action of a small external magnetic field when aligned as shown. This arrangement is known to have a theoretical Poisson's ratio of zero.[92] In fact the structure can be observed to contract laterally while keeping its length along the direction of the applied magnetic field. Deviation from the ideal behaviour was also observed. The reason for this is the relatively small number of repeating units which made edge effects significant. In fact, the outer parts showed a tendency to sheer. It should be further noted that when the direction of the external magnetic field was reversed the structure reconfigured itself as shown in Figure 5(e-h). The arrangement can be described as a mirror image of the one adopted with the original magnetic field. Made exception for this, its behaviour was the same for both orientations of the external magnetic field.

The analogous results for System II are illustrated in Figure 6. As it can be noted from this figure, when the external magnetic field was in one direction, the structure configured itself as a re-entrant hexagonal honeycomb (Figure 6(a-d)), while when the external magnetic field acted in the opposite direction it arranged itself as a conventional hexagonal honeycomb

(Figure 6(e-h)). In the former case, upon increasing the external magnetic field, the structure shrunk as expected, since this shape is known to have a negative Poisson's ratio. The opposite is true in the latter case with structure spreading out in both orthogonal directions. Again, this was expected since the conventional hexagonal honeycomb is known to have a positive Poisson's ratio. Yet, edge effects are visible, with some shearing clearly noticeable. The edge effects can be minimised by considering systems with a larger number of unit cells.

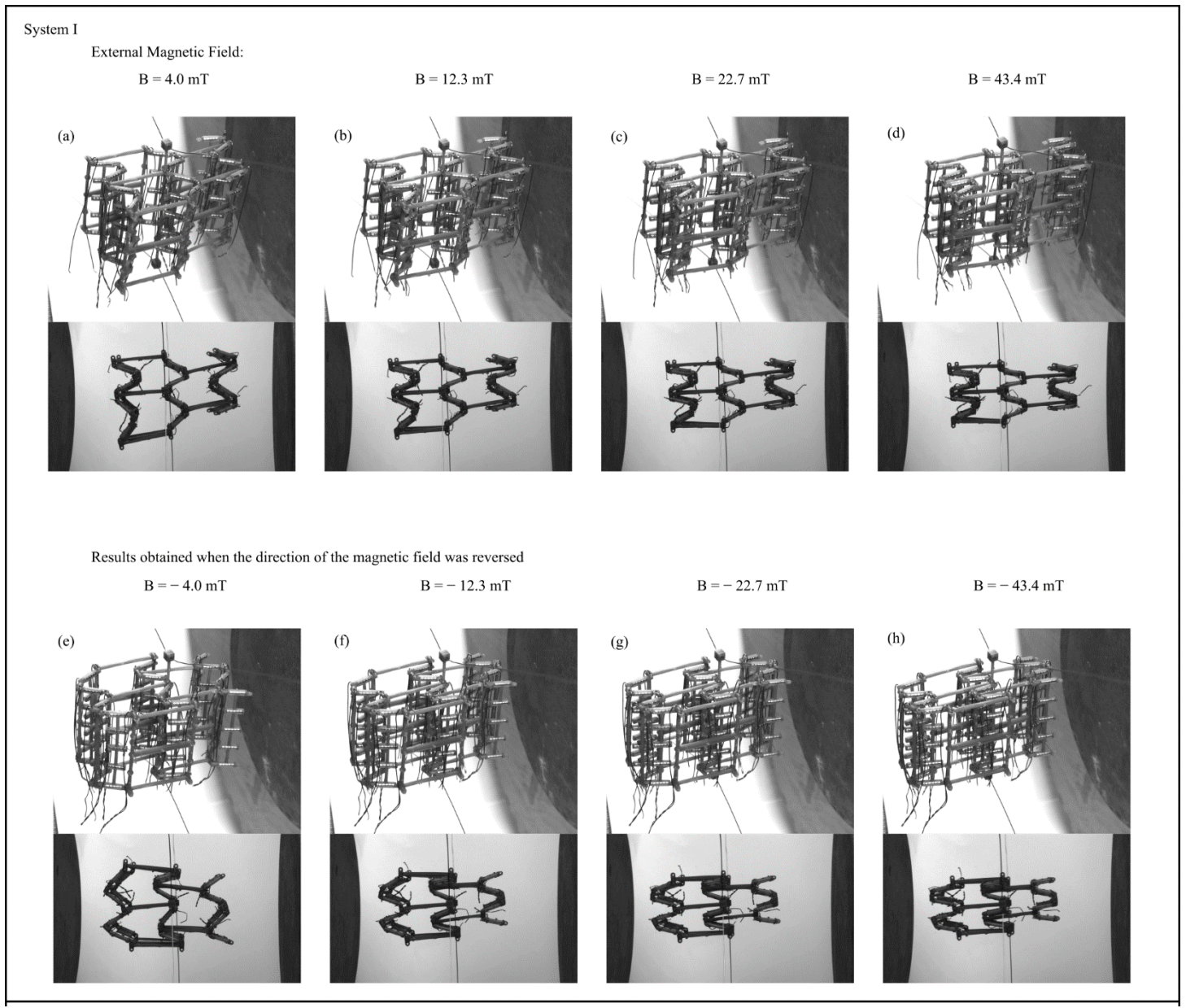

**Figure 5.** (a-h) Images from a side view and aerial view of a System I structure under an external magnetic field of 4.0, 12.3, 22.7, 43.4, −4.0, −12.3, −22.7 and −43.4 mT respectively. The negative sign indicated that the external magnetic field is acting in the opposite direction.

The results clearly indicate that the structure allows for detailed control of its dimension. In addition, it is possible to choose configurations that have negative, zero, or positive Poisson's ratio, with System II being able to flip between negative and positive Poisson's ratio arrangements simply by reversing the direction of the applied magnetic field.

Furthermore, additional control can be induced by replacing the emended permanent magnets with electromagnets. In this case it would be possible to selectively alter the magnetic moment locally inducing regional deformation effects.

All these properties allow the structures presented to be employable in a number of applications. Deployable systems, particularly in the aerospace industry is one of them. For example, they could serve as scaffolding for the solar wings in satellites. Here, they could be folded compactly for transport and then induced to open up during deployment. An alternative application would be for sound proofing systems where the configuration can be adjusted in real time in accordance with the wavelength of the sound travelling through the structure. Similarly, as a sieve it can be made to have adjustable pore sizes interactively, something that might find applications in the biomedical industry. The structures can also be designed to act as an actuator. In this case, nanoparticles can be used to produce miniature actuators. There is also the possibility of using the structure in robotics where local deformations can be used to induce locomotion. These are just a few of the possible applications that a system based on the proposed design can have.

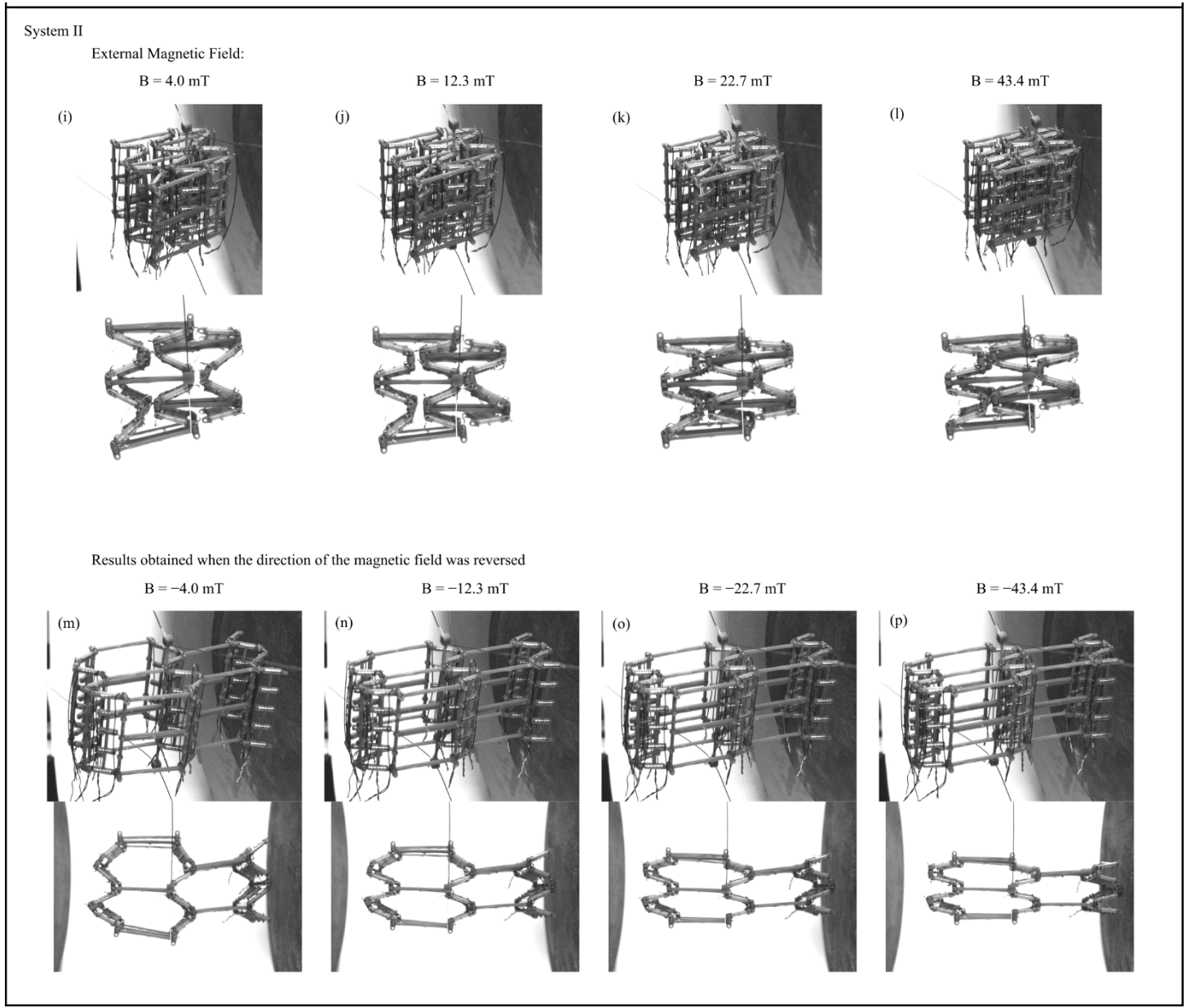

**Figure 6.** (a-h) Images from a side view and aerial view of a System II structure under an external magnetic field of 4.0, 12.3, 22.7, 43.4, −4.0, −12.3, −22.7 and −43.4 mT respectively. The negative sign indicated that the external magnetic field is acting in the opposite direction.

## 4. Conclusion

In this study, the magneto-mechanical metamaterial initially proposed by Galea et al.[33] was investigated further experimentally to determine how the magnetic moment of the emended magnets and the length of the arms affect the deformation of the structure when subject to an external magnetic field. Experimental prototypes having negative, zero, and positive Poisson's ratio were designed and manufactured by stacking the basic accordion-like structure in different ways. Experimental work indicated that the dimensions of all these configurations can be controlled by applying an external magnetic field. The theoretical studies by Galea et al.[33] were experimentally confirmed whereby it was shown that it is possible to flip between a negative and a positive Poisson's ratio configuration simply by reversing the direction of the applied magnetic field. The high degree of turnability and control that can be exerted on the structure, as well as the fact that their behaviour can be altered interactively, can potentially lead to a multitude of applications. Possible uses include, amongst others, the design of scaffoldings for deployable structures, actuators, variable pored sieves, and sound proofing systems. Such systems can find utilisation in different industries such as the aerospace, biomedical, and robotics ones.

It is envisioned that a potential application of the structures investigated in this work will be an untethered miniature multi-directional actuator that provides a high degree of real time control. The ability to modify the geometry of the structure interactively through the use of electromagnetics will allow it to be redeployed for different applications without the need of a physically distinct system. Furthermore, the actuator could be used to impart controlled movement in two directions simultaneously. In fact, the possibility of switching between positive and negative Poisson's ratio implies that the direction of the force exerted in the plane of deformation could be either in the same or in opposite directions. Given its high adaptability and controllability the actuator can be especially attractive in the aerospace, biomedical, and robotics industries. Apart from actuators, the properties exhibited by this structure can find potential use in deployable structures, including drones and satellites amongst others. In these cases, the system would be packed densely in the launching phase and subsequently deployed in situ through the use of magnetic fields. The process could allow the system to lock in the desired configuration without the use of pistons and other traditional mechanical parts. Real time adjustable variable pored sieves and sound proofing systems represent other possible applications of the structure. It is hoped that these encouraging results will lead to further investigations into these systems.

**Acknowledgements**

The research work disclosed in this publication was funded by the Tertiary Education Scholarship Scheme (Malta). It also uses equipment procured through a grant by the University of Malta for the project entitled 'A novel method to induce auxetic behaviour' (Grant number: GSCRP10-21). K.K.D. acknowledges the support of the Polish National Science Centre (NCN) in the form of the grant awarded as a part of the SONATINA 5 program, project No. 2021/40/C/ST5/00007 under the name "Programmable magneto-mechanical metamaterials guided by the magnetic field".

**Additional information:**

*Declaration of competing interest*: The authors declare that they have no known competing financial interests or personal relationships that could have appeared to influence the work reported in this paper.

*Figure colour*: All figures are meant to be printed in grey scale and available in colour only online.